\pdfoutput=1

\documentclass{mn2e}
\newcommand{\apropto}{\;
  \raise0.3ex\hbox{$\propto$\kern-0.75em\raise-1.1ex\hbox{$\sim$
  }}\;\hskip-2pt }
\usepackage[dvips]{graphicx}
\usepackage{subfigure}

\newcommand{\lta}{\;
  \raise0.3ex\hbox{$<$\kern-0.75em\raise-1.1ex\hbox{$\sim$
  }}\;\hskip-2pt }
\newcommand{\gta}{\;
  \raise0.3ex\hbox{$>$\kern-0.75em\raise-1.1ex\hbox{$\sim$
  }}\;\hskip-2pt }

\usepackage[cp1251]{inputenc}
\usepackage[T2A]{fontenc}
\usepackage[russian]{babel} 

\usepackage{color}
\usepackage[usenames,dvipsnames]{xcolor}

\title[ A new dynamo pattern revealed by the tilt angle of bipolar sunspot groups
]{ A new dynamo pattern revealed by the tilt angle of bipolar sunspot groups
 }

\author[A.~Tlatov, E.~Illarionov, D.~Sokoloff, V.~Pipin]
{A~Tlatov$^{1}$\thanks{email: tlatov@mail.ru},
E.~Illarionov$^2$\thanks{email: egor.mypost@gmail.com},
D.~Sokoloff$^{3}$\thanks{email: sokoloff.dd@gmail.com},
V.~Pipin$^{4}$\thanks{email: pip@iszf.irk.ru}\\
$^{1}${Kislovodsk Mountian Astronomical Station of the Pulkovo Observatory, 357700, Box-145, Kislovodsk, Russia}\\
$^2${Department of Mechanics and Mathematics, Moscow State University,
Moscow,119991, Russia}\\
$^{3}${Department of Physics, Moscow State University,
Moscow,119992, Russia}\\
$^4${Institute of Solar-Terrestrial Physics, Russian Academy of
Sciences, Irkutsk, 664033, Russia}\\  }
\begin{document}
\label{firstpage} \maketitle
\begin{abstract}
We obtain the latitude-time distribution of the averaged tilt 
angle of solar bipoles. For large bipoles, which are 
mainly bipolar sunspot groups, the spatially averaged tilt angle 
is positive in the Northern solar hemisphere and negative 
in the Southern, with modest variations during
course of the solar cycle. We consider the averaged 
tilt angle to be a tracer for a crucial element of the solar 
dynamo, i.e. the regeneration rate of poloidal large-scale 
magnetic field from toroidal. The value of the tilt obtained 
crudely corresponds to a regeneration 
factor corresponding to
about 10\% of r.m.s. velocity of solar convection. 
These results develop findings of Kosovichev and Stenflo 
(2012) concerning Joy's law, and agree with the usual 
expectations of solar dynamo theory. Quite surprisingly, 
we find a pronounced deviation from these properties 
for smaller bipoles, which are mainly solar ephemeral regions. 
They possess tilt angles of approximately the same 
absolute value, but of opposite sign compared to that of the large bipoles. 
Of course, the tilt data for small bipoles are less 
well determined than those for large bipoles; however they 
remain robust under various modifications of the data processing.

\end{abstract}
\begin{keywords}
Solar activity -- solar magnetic fields -- solar dynamo
 \end{keywords}

\section{Introduction}

There is a widely accepted idea that the solar magnetic activity, 
which is itself manifested in the quasiperiodic variation of sunspot 
number and other similar tracers, results from the dynamo waves 
of quasistationary magnetic field propagating in solar interior.
Differential rotation which stretches the toroidal magnetic field 
from poloidal is an obvious driver for the solar dynamo.
Differential rotation alone is not capable of exciting the dynamo. 
Additional drivers which can  create poloidal fields from 
toroidal ones are required. Symmetry arguments brought Parker (1955)
to the idea that the collective action of the cyclonic motions (associated
with the $\alpha$-effect) in the solar convection zone  can transform 
the toroidal component of the large-scale magnetic field to poloidal.
Steenbeck, Krause and R\"adler (see Krause and R\"adler, 1980)
provided the mathematical basis for this idea. It was found that 
the $\alpha$-effect results from the breaking of the reflection 
symmetry of the convective flows due to the global rotation. Currently, 
several different kinds of $\alpha$-effect  are widely used in solar 
dynamo models. The flux-transport and the Babcock-Leighton types of dynamo
employ a nonlocal $\alpha$-effect, which results from the Coriolis 
force acting on flux-tubes rising in the solar convection zone (see,
e.g.,  Choudhuri et al. 1995; Dikpati and Charbonneau 1999). They show that
this kind of dynamo can operate in the Sun in conjunction with 
meridional circulation. In mean-field dynamos the $\alpha$-effect  
works {\it in situ}. It is more then plausible (Dikpati and Gilman, 2001) 
that both effects contribute to solar dynamo action, and their relative 
importance differs between normal cycles and the Grand Minimum epochs. Anyway, 
the physical nature of the dynamo driver which connects toroidal magnetic 
field with poloidal remains an important problem in solar dynamo theory.

Helioseismology provides a substantial observational basis which can illuminate 
the properties of the solar rotation. The driver of the poloidal magnetic 
field in the solar dynamo is much less well known than the differential 
rotation which largely drives the toroidal field, 
because it is weaker  and requires more sophisticated 
observational schemes. Seehafer (1990) suggested that the  current 
helicity in solar active regions can be a tracer of the reflection 
symmetry breaking in the solar dynamo. It has  required about twenty 
years of intensive efforts to obtain a more or less clear message 
for dynamo studies from observations of the  current helicity observations 
(see e.g. Zhang et al. 2010). Currently, the nature of the the poloidal
magnetic field on the Sun remains a matter of intensive discussions.
In particular, solar dynamo models must assume fairly arbitrarily the 
value and form of the regeneration of poloidal 
magnetic field from toroidal, and the corresponding regeneration rate 
is a crucial governing parameter for the solar dynamo.
Therefore, there is a fundamental interest in getting observational information 
concerning the induction mechanisms  of the poloidal magnetic field 
in the solar dynamo, in order to learn which mechanism is most important.

An obvious relevant tracer here is the tilt angle of bipolar sunspot groups. 
As early as 1919 (Hale et al., 1919) Joy's law stated that that active 
sunspot regions tend to be "tilted"\,, with the leading spot closer to the 
equator than the following spot. This law gives, quite straightforwardly,
the desired link between toroidal and poloidal magnetic field. 
The importance of the Joy's law is that it gives the result of  the combined 
action of all sources contributing to the desired link between 
toroidal and poloidal fields, while the other tracers (say, kinetic helicity) 
illuminate contributions of particular mechanisms to the joint result. 

In spite of the fact that Joy's law is as old as the famous Hale 
polarity law, its impact on solar dynamo theory remained 
quite limited until recently, simply because the 
tilt angle data are much noisier than 
the sunspot polarity data. Recently, the accumulation of observational data 
and the progress of observational techniques, in particular due to the SOHO/MDI 
high resolution magnetograms, has provided  sufficient data. This can 
be used to suppress the noise by averaging, and gives convincing results 
concerning Joy's law. In our opinion, the breakthrough was achieved 
in the fundamental paper by Stenflo \& Kosovichev (2012), who convinced 
at least the solar dynamo community that the observations available allow 
the extraction of tilt angles with a reasonable level of confidence. Stenflo 
and Kosovichev (2012) presented  tilt angle as a function of solar 
latitude averaged over the solar cycle. In general, their results 
look quite consistent with expectations from solar dynamo theory. 

We believe that the available data enables us to make a further step 
and to clarify details of the temporal and latitudinal evolution of the 
tracers during the course of the solar cycle, as well as to learn the 
contribution in the tracer from bipolar sunspot groups of various 
intensities. The aim of this paper is to present the result of such 
an investigation, and to discuss it in the context of solar dynamos.

In presenting our results we exploit the standard methods in 
current helicity studies (Zhang et al. 2010), and obtain tilt angles
averaged over latitudinal and temporal bins overlaid on the sunspot 
butterfly diagram. According to the current naive understanding of solar 
dynamo theory we do not expect to see in such a diagram any very dramatic 
behaviour of the averaged tilt angle. In this sense we confirm and 
extend the conclusion of Stenflo \& Kosovichev  (as well as the 
earlier conclusion of Sivaraman et al. 1999) that the tilt angle 
does not vary dramatically during the course of the cycle. In contrast, we 
obtain, quite unexpectedly in the context of dynamo studies, 
that the tilt angle 
depends dramatically on the intensity of the bipolar group. 
Here we confirm and develop the preliminary conclusions of Tlatov and Obridko 
(2012). 
We present an interpretation of our results in the context of solar dynamo
modelling.

\section{Recognizing bipolar groups}

{We base our analysis on the magnetograms from Kitt Peak 
Vacuum telescope (KPVT), SOHO/MDI data (Scherrer et al, 1995; 
soi.stanford.edu/magnetic/Lev1.8/). In addition, we check the part of
the obtained results by using fits files of 720-second cadence 
HMI data (Schou et al., 2012).   Each magnetogram gives a map of
the line-of-sight component of the magnetic flux density averaged over
the spatial resolution window. We skipped from our analysis the magnetograms that have defected pixels (e.g., after the solar
flares). The magnetograms that cover only part of the solar disk were  skipped, as well.}

Similar to Tlatov et al. (2010) and, also,   Stenflo \& Kosovichev (2012), 
we use a computer algorithm 
to isolate bipolar regions in solar magnetograms. The algorithm we use 
was suggested by Tlatov et al. (2010) and allows isolation of 
bipolar groups with low surface area or magnetic flux. This is why our 
data sample is richer than that analyzed by Stenflo \& Kosovichev (2012),
and it is essential in getting a tilt butterfly diagram. 
Stenflo \& Kosovichev (2012) 
isolated about $1.6 \times 10^5$ bipoles from about $7.4 \times 10^4$ 
magnetograms, i.e. about 14 magnetograms per day, and so a rate 
of  about 2 bipoles per day.

They considered active regions only. Because the lifetime of active 
regions substantially exceeds 1 day, the tilt information is
redundant in that the tilt of a given bipole is measured many times. 
Our database for the same period 1996-2011 (MDI data) contains about
$6.8 \times 10^4$ bipoles from about $4.5 \times 10^3$ magnetograms, 
i.e. 1 magnetogram per day,  and so about 15 bipoles per day. For the period 
1975-2003 (KPVT data -- 3 cycles) 
we obtained about $2.45 \times 10^5$ bipoles. 
{
To verify that these samples are not affected
by noise or other known issues of MDI data (Liu and Norton, 2001) 
we used additionally HMI data (2010-2012), from which
we obtained about $9.7 \times 10^3$ bipoles (1 magnetogram per day).
}
Apart from sunspot groups, our bipoles represent small ephemeral regions.

{For the sake of consistency we recall here  the basic 
elements of the algorithm suggested by Tlatov et al. (2010). 
The image with an example of the recognized bipolar groups can be
found in the above cited paper}

{In Step 1, we look through the data set a for each magnetogram,  
we sift defective pixels do one nearest-neighbor 
smoothing (see Tlatov et al., 2010).  Then,  we select the monopole domains with 
magnetic field exceeding the threshold level $B_{\rm min}$ for positive
and negative values of $B_{\rm min}$ separately. 
For each connected domain the following parameters are then
calculated: the heliographic coordinates of the geometrical center of the domain
including the projection effect, its area measured in millionths of 
the solar hemisphere (MSH\footnote{
1 MSH = $3.044 \times 10^{6}$ $\rm km^2$.
The round shaped spot with area S (in MSH) has a diameter  
$d=1969\sqrt{S}$ $\rm km$ = $(0.1621\sqrt{S})^{\circ}$.
See Vitinsky et al., 1986.}), and the magnetic flux through this area.
}

{ The next stage is to search for a domain of given polarity (domain 1) 
and a corresponding domain (domain 2) of the opposite polarity. We sift the structures
with area less than 20 MSH from the search
The search is performed in a circular region whose center is located 
at the geometrical center of the initial domain and with radius
$7^\circ+dL/2$ on the solar disc, where $dL$ is a longitudinal size
of this domain. Note that the value of $7^\circ$ corresponds to the
mean distance between pores in bipolar sunspot groups of 
the Z\"urich Class B (Vitinsky et al., 1986).
It looks reasonable to introduce a criterion that both domains have 
more or less equal magnetic fluxes, i.e. $||F_1|-|F_2||/(|F_1|+|F_2|)<\varepsilon$,
and to require that that the distance between the corresponding 
domains is minimal. Each search is started with
an arbitrary pole of positive polarity. For this element the program
finds the best neighbor match for the negative polarity pole. 
After this, a reverse search is performed to find
the closest neighbor for the negative pole. If the search returns
the same pair of negative and positive poles, the pair is marked
and is taken out of the next search.
The search was restricted to the central zone, 
within $\pm 60^{\circ}$ from the central meridian of solar disc.
}

{ Finally, we calculate the tilt angle $\mu$ as the angle between a line 
connecting the geometrical centre  of the domain with the negative polarity 
with the centre of the domain with positive polarity  in a
bipolar group, and the solar equator. 
We associate the tilt with the bisecting point of this line.
If the leading polarity is positive 
the tilt is close to $0^\circ$ while if the leading polarity is negative the
tilt is near $180^\circ$. 
}

\begin{figure*}
\centering
\includegraphics[width=0.68\linewidth]{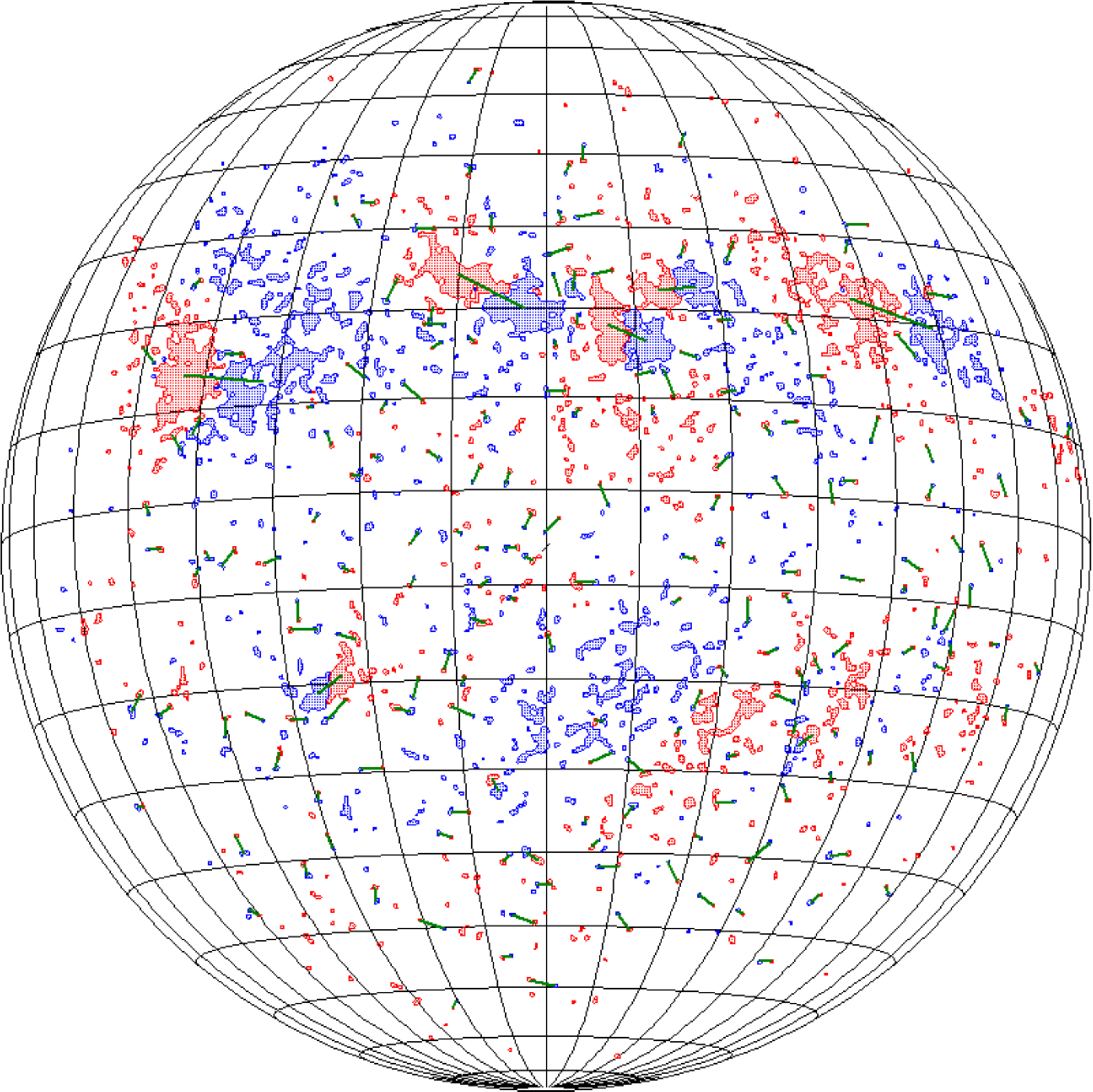}
\caption{Examples of bipole identification at MDI magnetogram
from 2011 April 11.
}
 \label{Bip}
\end{figure*}
{
Tlatov et al. (2010) demonstrated that the algorithm is quite
robust. Examples of the identified bipoles are given in Fig.~\ref{Bip}.
Of course, it can sometimes give identifications that look irrelevant,
especially for small bipoles. However the main statistical properties
seem reliable to within a general level of noise. Basically we used one
magnetogram per day, but for the MDI data set we made an additional
check with 7 magnetogram per day and find the same results (within the
level of noise). The given results were confirmed by the procedure
where the search domain was restricted to the central zone within $\pm
30^{\circ}$ from the central meridian
of the solar disc. These support the conclusion about robustness of the
principal algorithm.

As opposed to Stenflo \& Kosovichev (2012) we used significantly lower
smoothing and lower values of threshold levels for magnetic field.
We obtained individual substantive
domains with appropriate coordinates and fluxes.
It allow us to perform an analysis for substantially smaller structures
versus Stenflo \& Kosovichev (2012). There are also some differences in
determination of tilt angle. We calculate tilt between two domains with
similar fluxes rather then between geometric averaged fields with
different directions that usually consist of many unconnected regions
inside the box used by Stenflo \& Kosovichev (2012).
}

Additional information obtained during the identification of the groups allows 
separation of groups of given ranges of areas. 

Obviously, it is useful for the analysis to obtain as many 
bipoles as possible. On the other hand, as more bipoles are collected
weaker bipoles have to be included in the database.  We recognize 
that it is difficult to determine mean tilt
 for weak bipoles, because  
the tilt angle distribution is almost homogeneous.
We expect, and confirm the expectation by the following 
analysis, that some features of bipole distribution become unstable 
if the weakest bipoles are included. Therefore we 
set a lower limit on the area of bipoles - they should exceed 50 MSH.
Smaller bipoles require more sophisticated analysis. Below we grade
the results obtained with respect to the level of their stability 
according to various selection criteria applied. Some statistical 
data concerning database used are given in Table 1.

\begin{table}
\begin{tabular}{llll}
\hline
Selection criteria & KPVT & MDI & HMI\cr
\hline
$50<S<300$ MHS & 201 221 & 50 195 & 6754 \cr
$S> 300$ MHS & 44 821  &  18 096 & 2911 \cr
\hline 
\end{tabular}
\caption{The number of bipoles in the database investigated}
\end{table}

\begin{figure}
\centering
\includegraphics[width=0.9\linewidth]{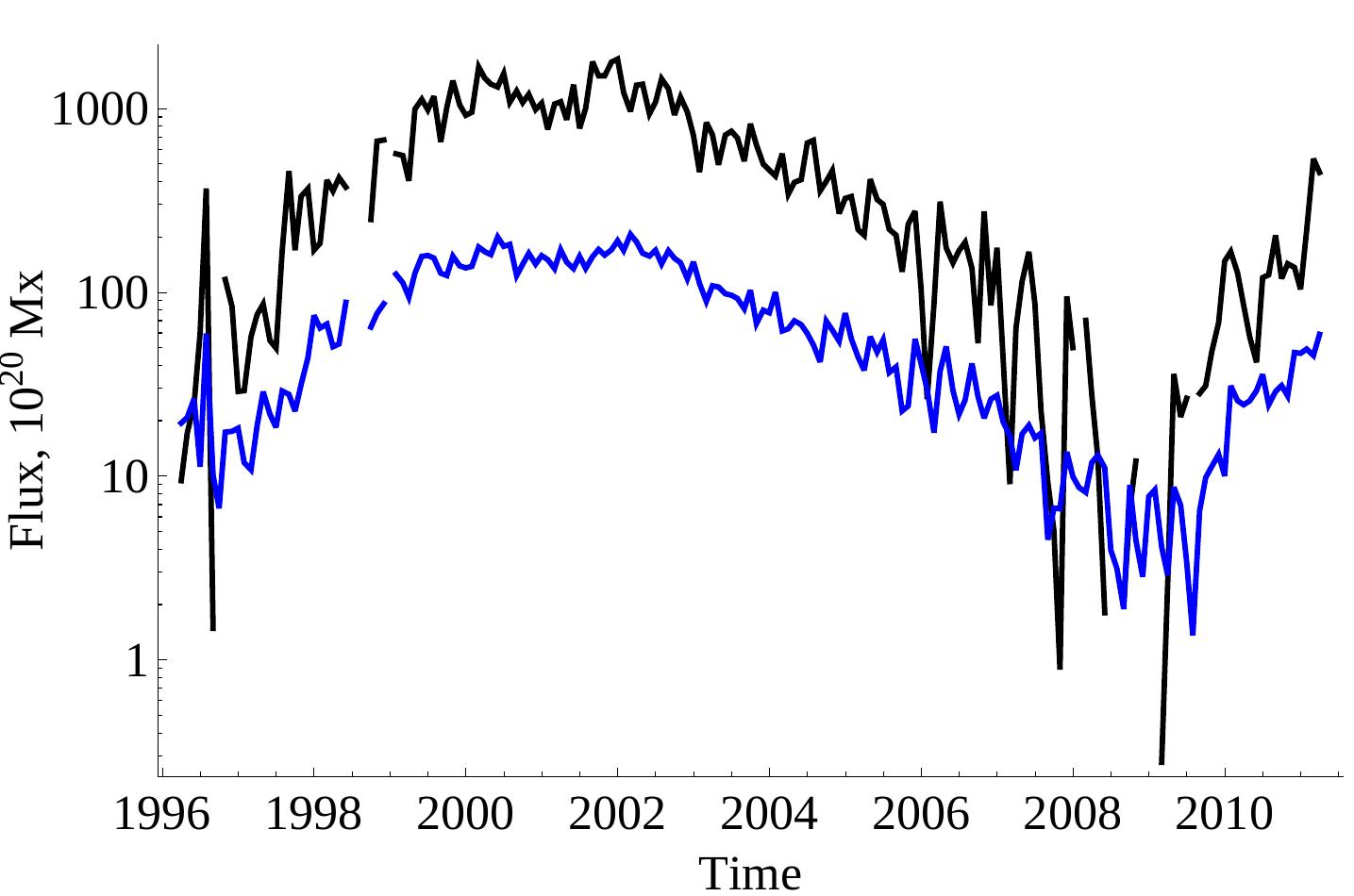}
\caption{The flux according to MDI data. The black line corresponds to areas
$S>300$ MSH, blue line shows the total flux from bipoles with areas $50<S<300$ MSH.
}
 \label{flux}
\end{figure}

Fig.~\ref{flux} presents graphics of total flux according to MDI data
for small ($50<S<300$ MSH) and large ($S>300$ MSH) bipolar groups. 
The lower bound $S=50$ MSH is chosen because the tilt angle distribution 
becomes practically homogeneous for smaller bipoles. The bounding value 
$S=300$ MSH is chosen because the tilt angle properties are more 
or less stable for larger bipoles. Another important point here is 
that the size of such bipoles more or less corresponds to the 
super-granulation scale. We see that both small and large bipoles are involved 
in the solar cycle. Of course, the contribution 
from small bipoles to the total magnetic flux is substantially smaller 
than that from the large bipoles.

\begin{figure}
\centering
\includegraphics[width=0.95\linewidth]{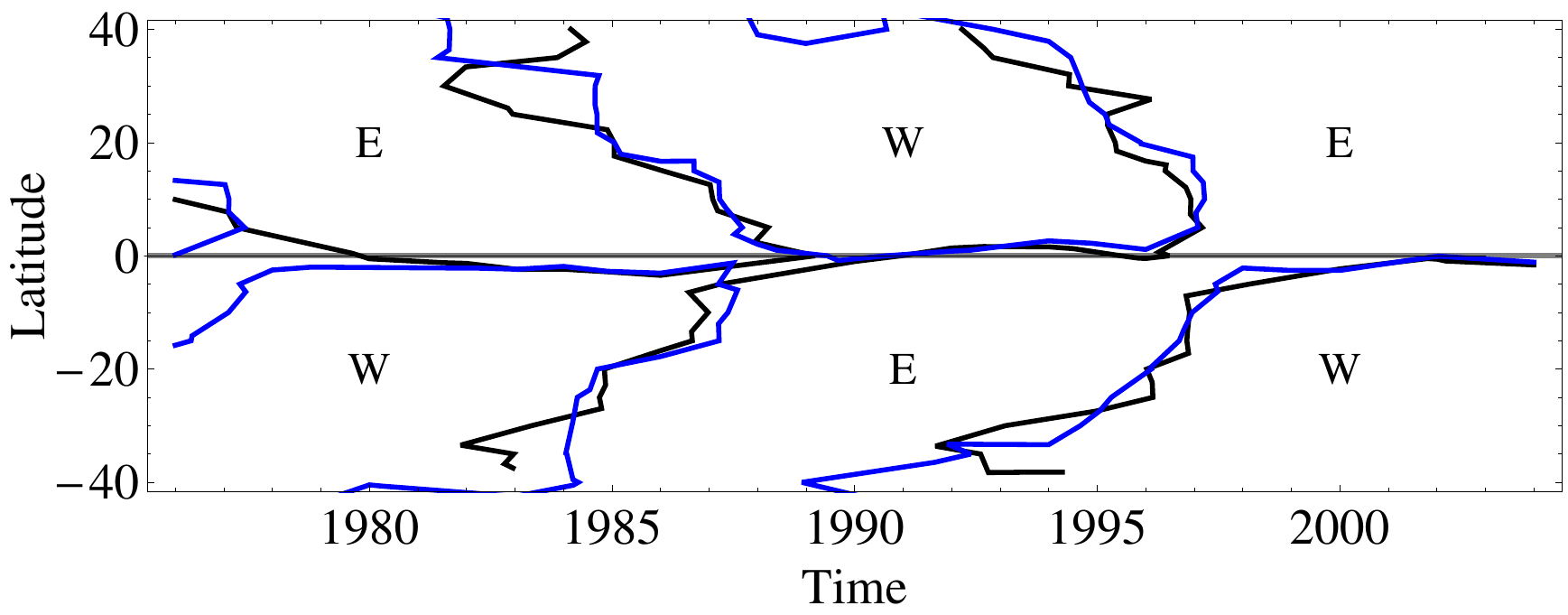}
\caption{Illustrating the prevalent E-W direction of tilt according to KPVT. 
Black shows areas $S>300$ MSH, blue the areas $50<S<300$ MSH. 
$B_{\rm min} = 10$ G.}
 \label{Fig8a}
\end{figure}

We demonstrate the correctness
of our selection of small and large bipolar groups as follows. 
According to the Hale polarity law the sunspot groups have 
opposite leading polarity in successive cycles in one hemisphere,
as well as between hemispheres in a single cycle. Fig.~\ref{Fig8a} 
confirms this symmetry for small bipoles and shows that 
the wings of the butterfly diagram which separate regions of 
the opposite sign almost coincide for large and small 
bipoles: black lines separate areas with prevalent
number of large bipolar groups ($S>300$ MSH) in east or west direction. 
The blue lines separate the same areas for smaller bipolar groups 
($50<S<300$ MSH). We consider the fact that the areas separated 
by blue lines are quite similar to the areas separated by black 
lines as a confirmation of correctness of our selection.

\section{Averaging tilt}

Obviously, the tilt for a given bipolar region is a quite noisy 
quantity and we are interested in the tilt data averaged over 
suitable latitudinal and temporal intervals.

\begin{figure}
\centering
\includegraphics[width=0.9\linewidth]{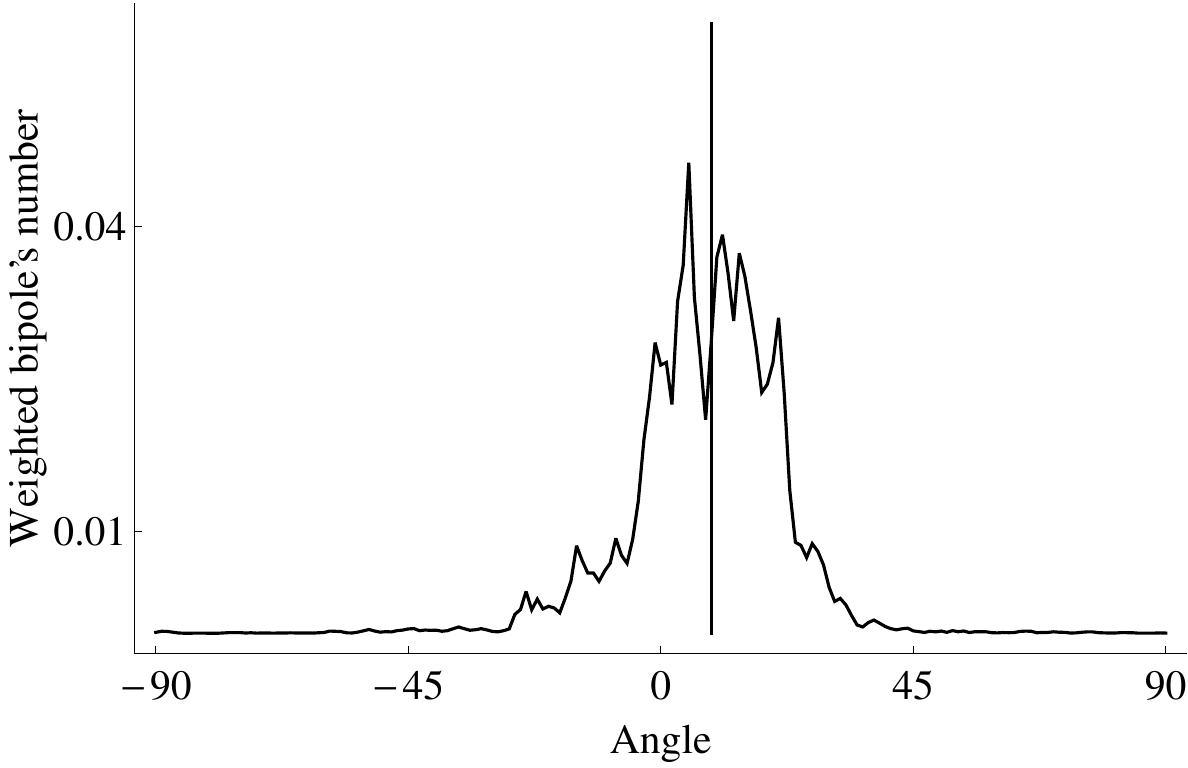}
\caption{An example of the tilt distribution for the MDI data (for 2001 - 2002) at 
$12.5^\circ\le\theta\le17.5^\circ$ for areas $S>300$ MSH (weighted data). 
The vertical line indicates the median of sample. 
Graphic is normalized to unit square.}
 \label{dist}
\end{figure}

For averaging we use 2-year time bins and $5^\circ$ latitudinal 
bins and then consider the data for a given activity cycle as required.
An example of the tilt angle distribution for a given time-latitude bin 
is shown in Fig.~\ref{dist}.

In averaging the distribution we have to take into account that the method
applied to measure the tilt can give rare and strong non-Gaussian 
deviations from the mean. For example, a bipolar group with inverse
polarity which violates the Hale polarity law gives by definition 
a derived tilt which differs from its true value by $180^\circ$. 
This happens only
in about 5\% of cases (e.g. Sokoloff and Khlystova, 2010).
The relative number of small bipoles that violate the polarity law 
is larger than that  for sunspot groups, about 10--20\% and they 
quite usually give an additional maximum in the tilt angle distribution, 
displaced from the primary by $180^\circ.$
However the tilt is usually quite close to $0^\circ$ or $180^\circ$ 
and we discard rare and strong non-Gaussian deviations.

We perform this averaging as follows. In the first step, we select for a given 
bin a hemisphere, either East (tilts $-90^\circ < \mu < +90^\circ$) or West 
(tilts $90^\circ< \mu < 270^\circ$), from where the tilt data are taken 
for averaging. We approximate the tilt distribution in each 
hemisphere by the Gaussian $\frac{A}{\sigma}\exp(-(x-a)^2/\sigma^2)$,
where $A$ is an amplitude, $a$ is a mean, and $\sigma$ is a standard 
deviation, and choose the hemisphere where $A$ is larger. As a rule 
this means that the number of tilts which are directed in this hemisphere 
is larger than directed in the opposite hemisphere. 
If desirable, we can weight the data using, say, the areas of the bipoles. 

In the second step we apply the robust statistical measure (median) to estimate 
the desired mean value and corresponding uncertainties. 
We order the tilt angles from the bin as  $\mu_1\leq\mu_2\leq...\leq\mu_n$,
$i=1...n$ and take $\mu_{((n+1)/2)}$ as the averaged quantity  
$m$ for the tilt if $n$ is odd and $\frac{1}{2}(\mu_{(n/2)}+\,\mu_{(n/2+1)})$, 
if $n$ is even. 
For large $n$ the estimate becomes Gaussian with  mean $m$ and standard deviation 
$\sqrt{\pi\frac{1}{n-1}\sum{(\mu_i-m)^2}}$ and then we can estimate
uncertainties using the standard Student criterion with $n-1$ degree of 
freedom (see Huber 1981).

Note that the number of large bipoles in our sample decreases sharply 
with the growth of bipole area. In order to increase the contribution of
the largest bipoles (including sunspots) to the average we use
weighting. Small bipoles largely determine the non-weighted averaging. 
We define the tilt as being positive if it is clockwise 
and negative if counterclockwise, regardless of hemisphere.

\section{Results}
\subsection{Joy's law}

We start our analysis by verifying the results of Stenflo \& Kosovichev
(2012) for our sample, assuming that the tilts scale with co-latitude 
as $k\sin \theta$.

\begin{figure}
\centering
\includegraphics[width=0.9\linewidth]{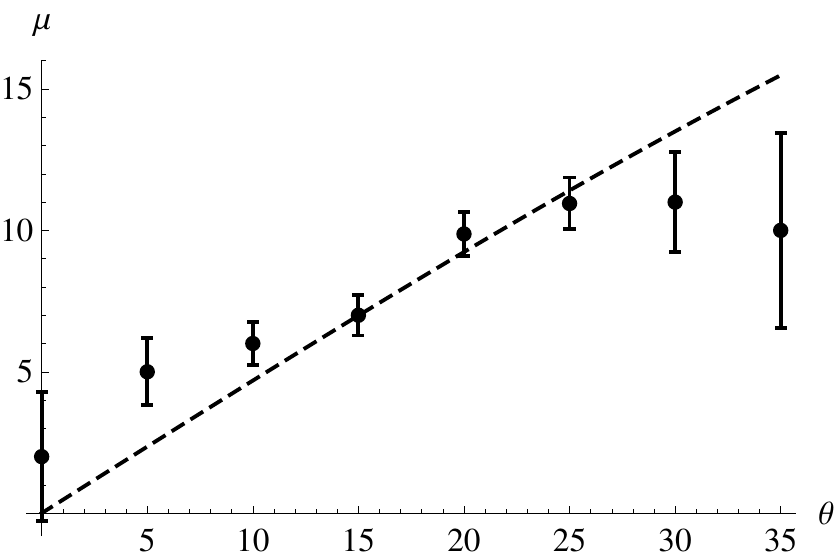}
\caption{Joy's law for MDI data, bipoles with the minimal magnetic field limit 
$B_{\rm min}=10$ G, area $S>300$ MSH. The error bars indicate 95\% confidence
interval. The dashed curve is the fit function
$\mu=27\sin\pi\theta/180^\circ$.  }
 \label{Fig1}
\end{figure}

In Fig.~\ref{Fig1} we present the averaged tilts for the  MDI data for 1996--2011
with combined northern and southern hemispheres, following the presentation
of the data by Stenflo \& Kosovichev (2012).
We selected the bipoles with the minimal magnetic field limit $B_{\rm min}=10$ G 
and took those with area $>300$ MSH. The data are weighted according 
to the bipole areas. 
We see from Fig.~\ref{Fig1} that Joy's law is in the form 
discussed by Stenflo \& Kosovichev (2012), except for the smaller inclination
of the fitted curve because of the bending of Joy's law at $30^\circ$ 
(the bending is absent in Stenflo \& Kosovichev, 2012 plots and 
the nature of the bending deserves a further clarification).

\begin{figure}
\centering
\includegraphics[width=0.9\linewidth]{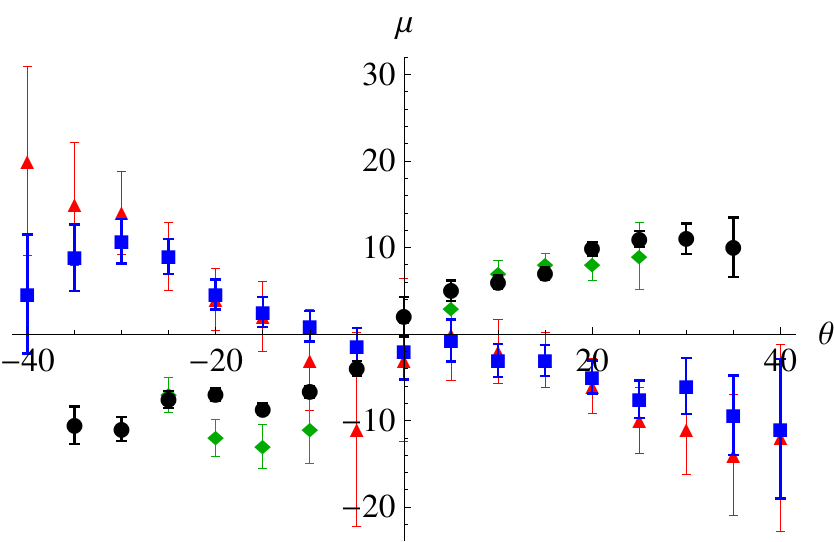}
\caption{
Joy's law according to MDI and HMI data. Black squares correspond to
areas $S>300$ MSH for MDI, green diamonds -- to
areas $S>300$ MSH for HMI. Blue points correspond to areas 
$50 < S <300$ MSH for MDI, red triangles -- to areas $20 < S <100$ MSH for HMI.
$B_{\rm min} = 10$ G for MDI and 15 G for HMI.}
 \label{Fig2}
\end{figure}

The correlation changes dramatically when smaller bipoles are taken into account.
We can see from Fig.~\ref{Fig2} that bipoles with areas 50-300 MSH
possess the opposite sign of tilt. This average for small bipoles is
not weighted. Note that, as for the large bipoles, there is a slight bending
in Joy's law for small bipoles at latitude  $30^\circ-40^\circ$.

{ We obtain that the analysis of HMI data confirms 
the above obtained unusual behaviour of small bipoles.  
The short period covered with HMI (2010-2012) does not allow us to 
make detailed estimations, but at qualitative level we can see 
from Fig.~\ref{Fig2} that small bipoles also have the inverse Joy's law.
The thresholds $B_{\rm min}$ and $S$ for HMI were shifted 
to keep daily number of bipoles comparable to MDI sample over a period
of its simultaneous work (2010-2011).
}

\begin{figure}
\centering
\includegraphics[width=0.9\linewidth]{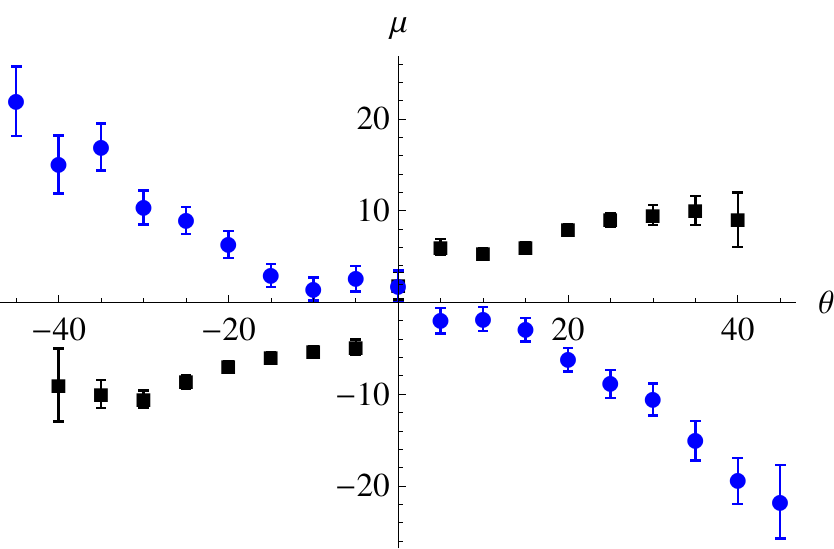}
\caption{Joy's law for KPVT, $B_{\rm min} = 10$ G; black squares correspond to
areas $S>300$ MSH, blue points correspond to $50< S <300$ MSH.}
 \label{Fig3}
\end{figure}

We verify the above conclusions about the bipoles from a longer period
of observation (KPVT data, 1975-2003, i.e.3 cycles), 
$B_{\rm min} = 10$ G (Fig.~\ref{Fig3}). 
We see from Figs.~\ref{Fig3} and ~\ref{Fig2} that the absolute value 
of tilt for small bipoles rises continuously from the equator to middle latitudes
(inverse Joy's law).

\begin{figure}
\centering
\includegraphics[width=0.9\linewidth]{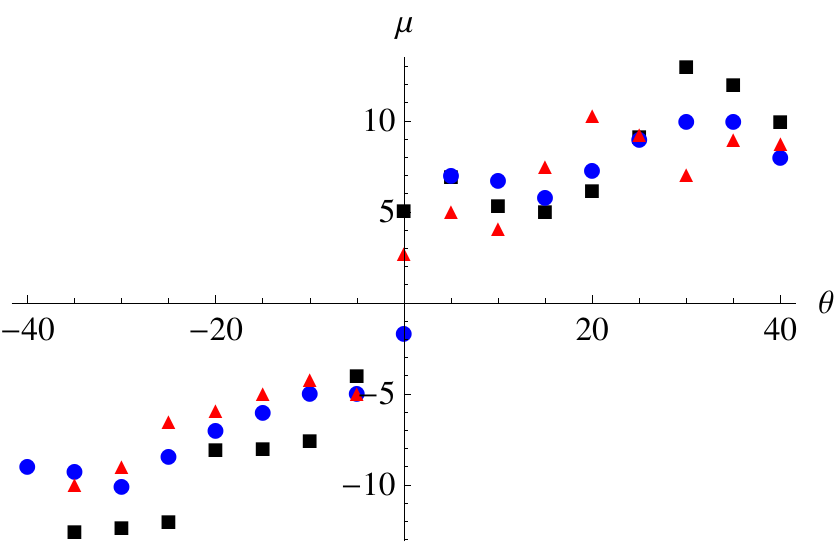}
\caption{Joy's law according to the weighted KPVT data, $B_{\rm min} = 10$ G, 
area $S>300$ MSH. 
Black squares - cycle 21, blue circles - cycle 22, red triangles - cycle 23.}
 \label{Fig4}
\end{figure}

Note that the available data
allow derivation of the correlations for cycles 21, 22 and 23 separately 
and confirm that the value $k$ does not vary substantially from one 
cycle to the next, see Fig.~\ref{Fig4}. 
{We do not observe such variations for cycles 23 and 24 
as well (Fig.~\ref{Fig2}).}
Our interpretation 
is that the weighting suppresses efficiently the impact of small bipoles.

\begin{figure}
\centering
\includegraphics[width=0.9\linewidth]{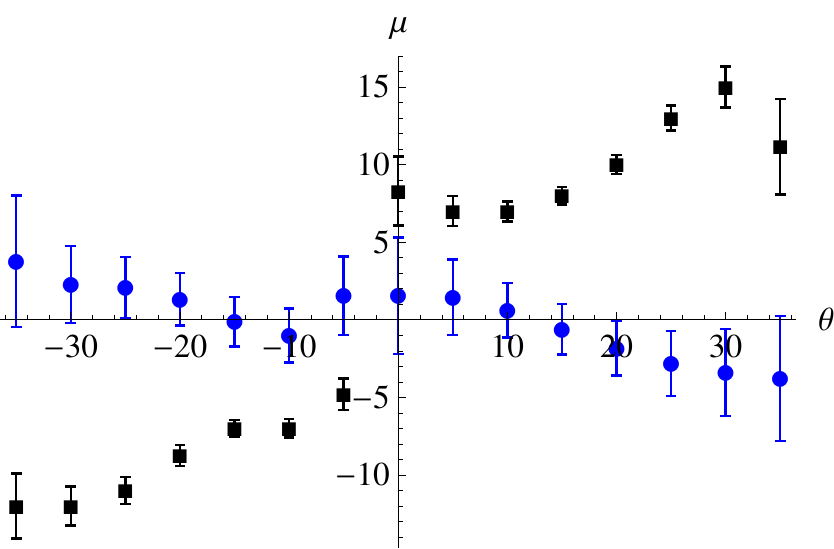}
\caption{Joy's law for KPVT, $B_{\rm min} = 100$ G, black squares correspond to
areas >100 MSH (weighted tilts), blue points to areas $S<100$ MSH
(non-weighted tilts).}
 \label{Fig5}
\end{figure}

Obviously, tilt can depend not only on area, but also on the strength of the
magnetic field. To check this we made a selection taking $B_{\rm min} = 100$ G
for the KPVT data. Fig.~\ref{Fig5} shows that tilts for small bipoles (<100 MSH)
have the opposite sign, but with lower value than for the larger bipoles.

\subsection{Butterfly diagram: tilt data and sunspot data} 

We move now to the space-time distribution of the averaged tilt angle
data overlaid on the sunspot butterfly diagram. As we learn from 
the above analysis of Joy's law, we have to consider contributions from
large and small bipoles separately, using an appropriate 
selection and weighting procedure.

\begin{figure*}
\centering
\includegraphics[width=0.9\linewidth]{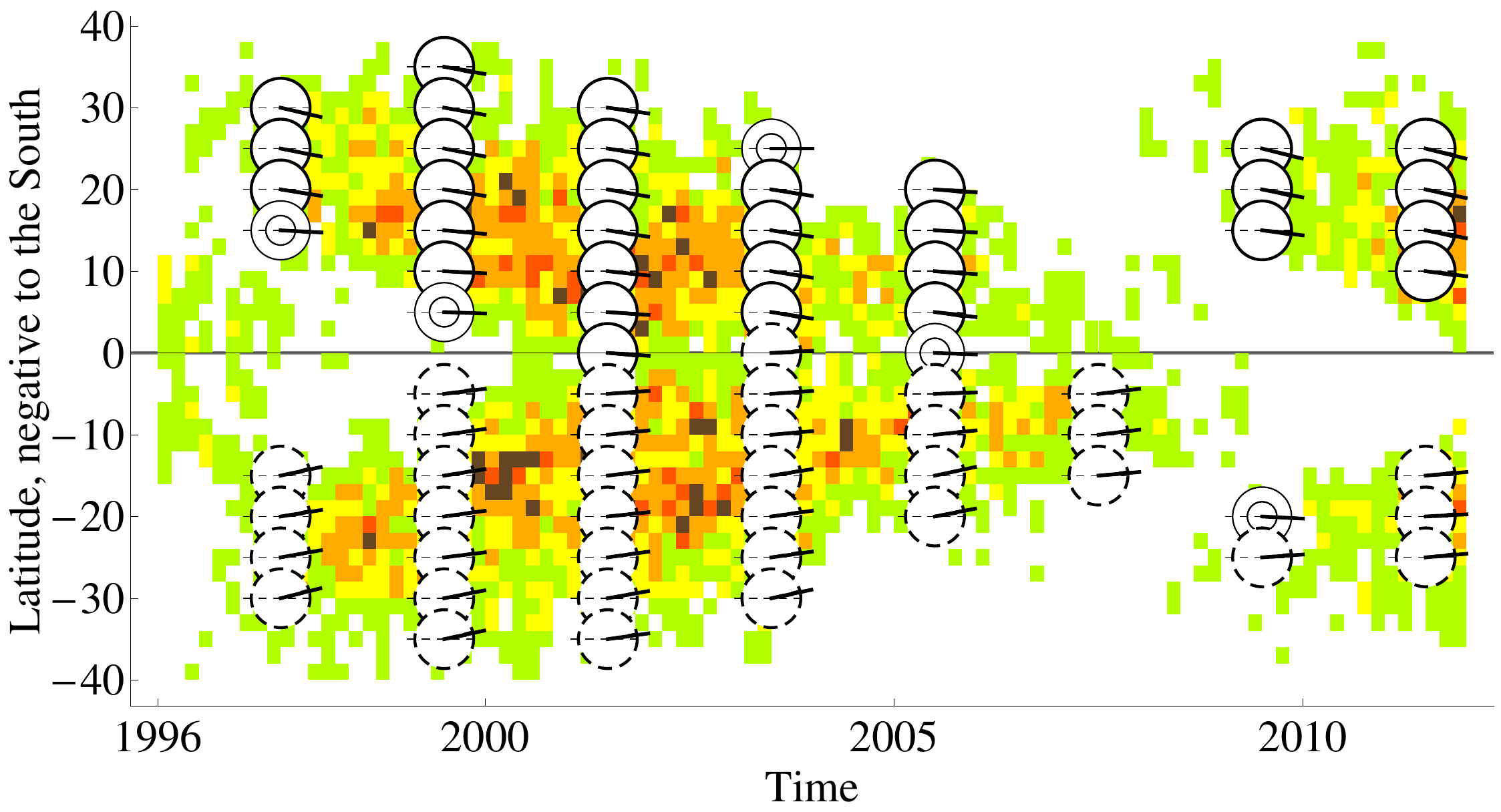}
\caption{Showing the butterfly diagram (tilt data overlaid sunspot data) 
for cycle 23, from weighted MDI data, $B_{\rm min}=10$ G, areas $S>300$ MSH.  
The tilt angles are shown as circles. Solid circles correspond to the bins 
with positive mean value of the tilt angle and zero tilt does not
belong to the confidence intervals for this mean value. Dashed circles 
correspond to the bins with negative
mean value of the mean value of the tilt angle and zero tilt does not 
belong to the confidence intervals for this mean value.
Double circles stand for the bins where zero tilt belongs to the confidence
intervals. Note that tilt is positive if the tilt is clockwise and negative 
if the tilt is counterclockwise regardless of hemisphere. }
 \label{Fig6}
\end{figure*}

Fig.~\ref{Fig6} shows the distribution for the MDI data for sunspots, 
cycle 23. 
We see from the Figure that an averaged positive tilt is strongly preferred 
in the Northern hemisphere, while negative tilt is strongly preferred in
the Southern.  A few minor exceptions from this polarity rule can be 
seen at the very end of the cycle and  near the solar equator. 
Exceptions for a corresponding polarity rule are
known as well for the current helicity data, e.g. Zhang et al. (2010).

\begin{figure*}
\centering
\includegraphics[width=0.9\linewidth]{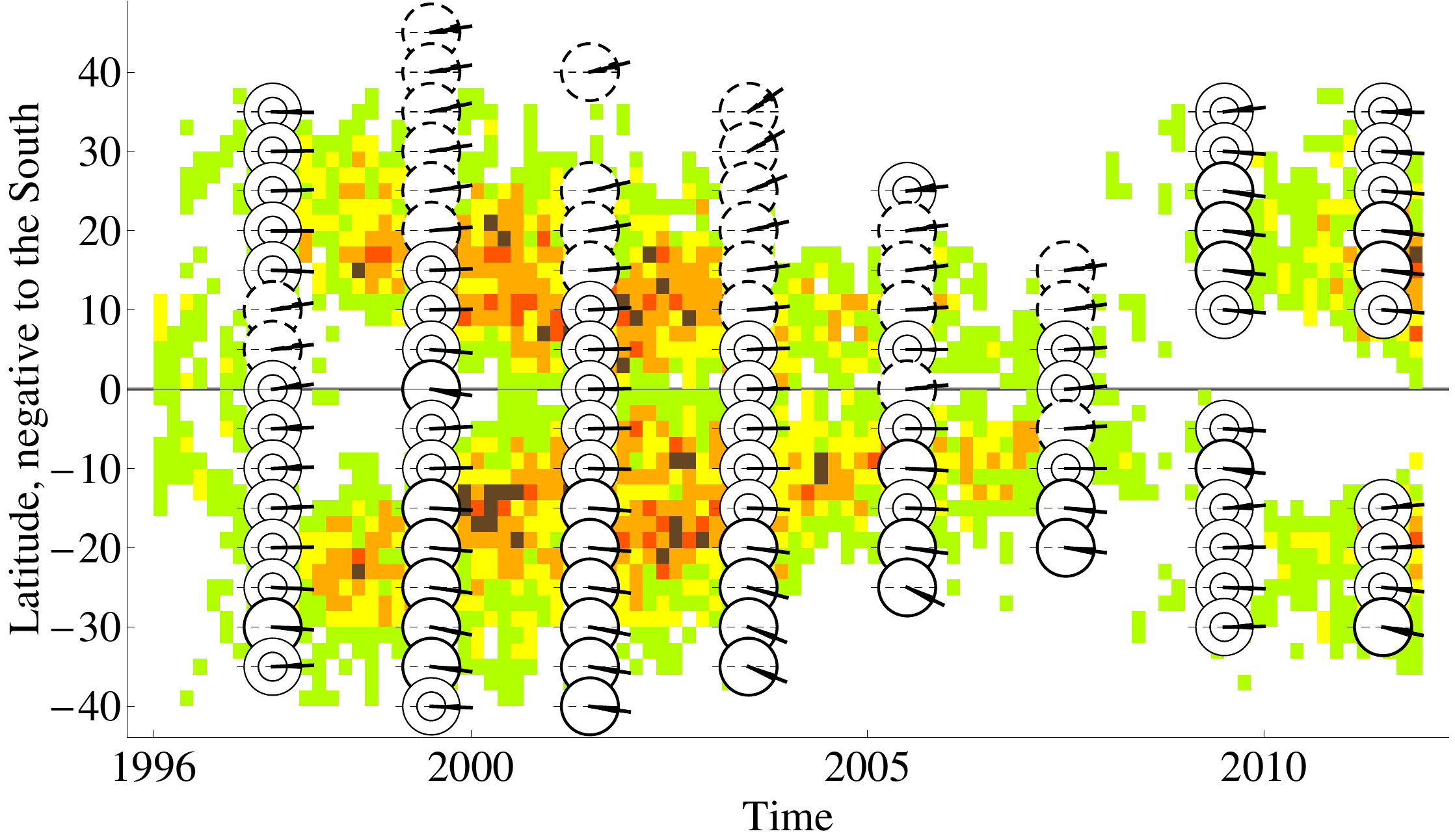}
\caption{Butterfly diagram (tilt data overlay sunspot data) for cycle 23, 
non-weighted MDI data, $B_{\rm min}=10$ G, areas $50<S<300$ MSH. Notation as
in the previous figure.}
 \label{Fig7}
\end{figure*}

Fig.~\ref{Fig7} shows that the polarity rule for the tilt is reversed 
if we consider only the contribution 
of the small bipoles. 

\subsection{Butterfly diagram: tilt data}

\begin{figure}
\centering
\includegraphics[width=0.95\linewidth]{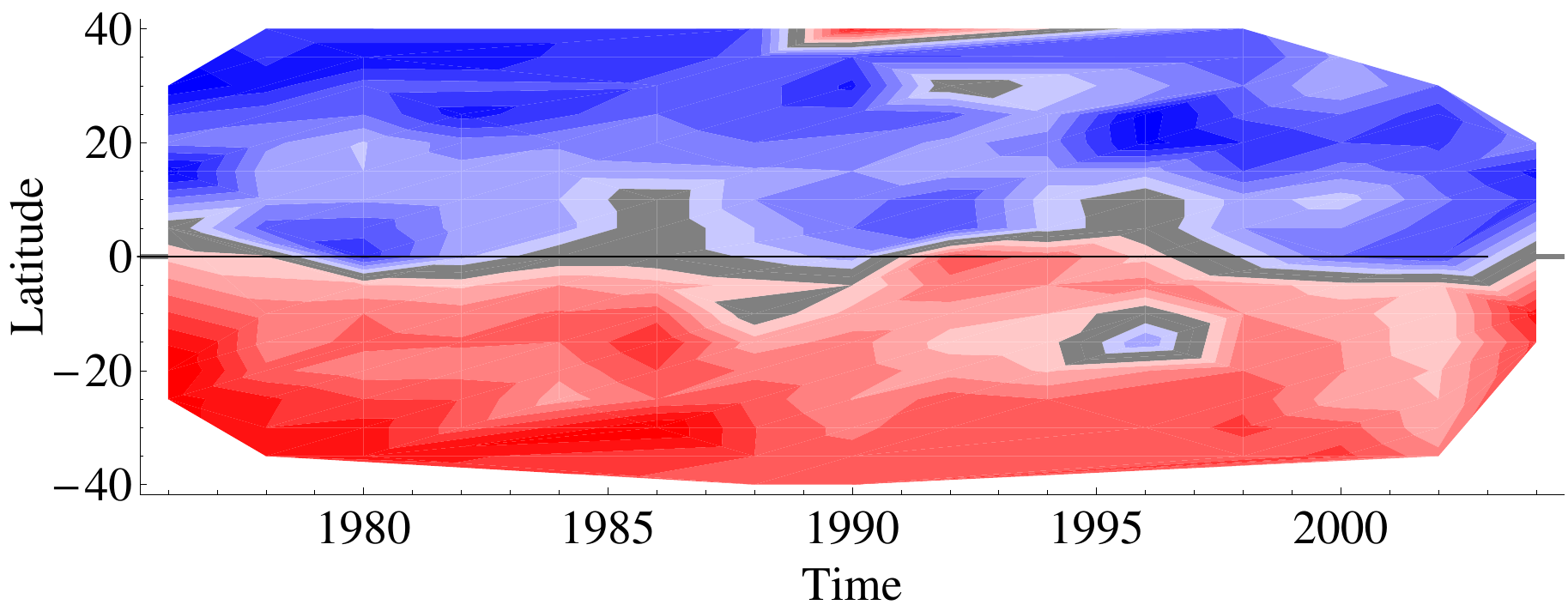}
\caption{Butterfly diagram for tilt alone according to  KPVT. Blue is 
for positive tilt, red is for negative tilt, grey is used to indicate 
areas where the tilt is zero to within the confidence intervals. 
$B_{\rm min} = 10$ G, areas $S>300$ MSH. \label{Fig8}}
\end{figure}

The next figures present the time-latitude distribution of the tilt data 
alone by colours: blue means positive tilt and red negative tilt; whereas
gray means that the tilt is  zero to within the confidence intervals. We see 
in Fig.~\ref{Fig8} that the tilt is mainly positive in the North hemisphere 
and mainly negative in the South hemisphere. Some exceptions from this 
polarity rule are associated mainly with the areas between cycles. 
Equatorward propagating patterns are weak but still visible inside the cycles. 
In places they have a double-peak structure: maxima are located at 
the beginning of the cycle at higher latitudes and at the end of 
the cycle at lower latitudes. 

\begin{figure}
\centering
\includegraphics[width=0.95\linewidth]{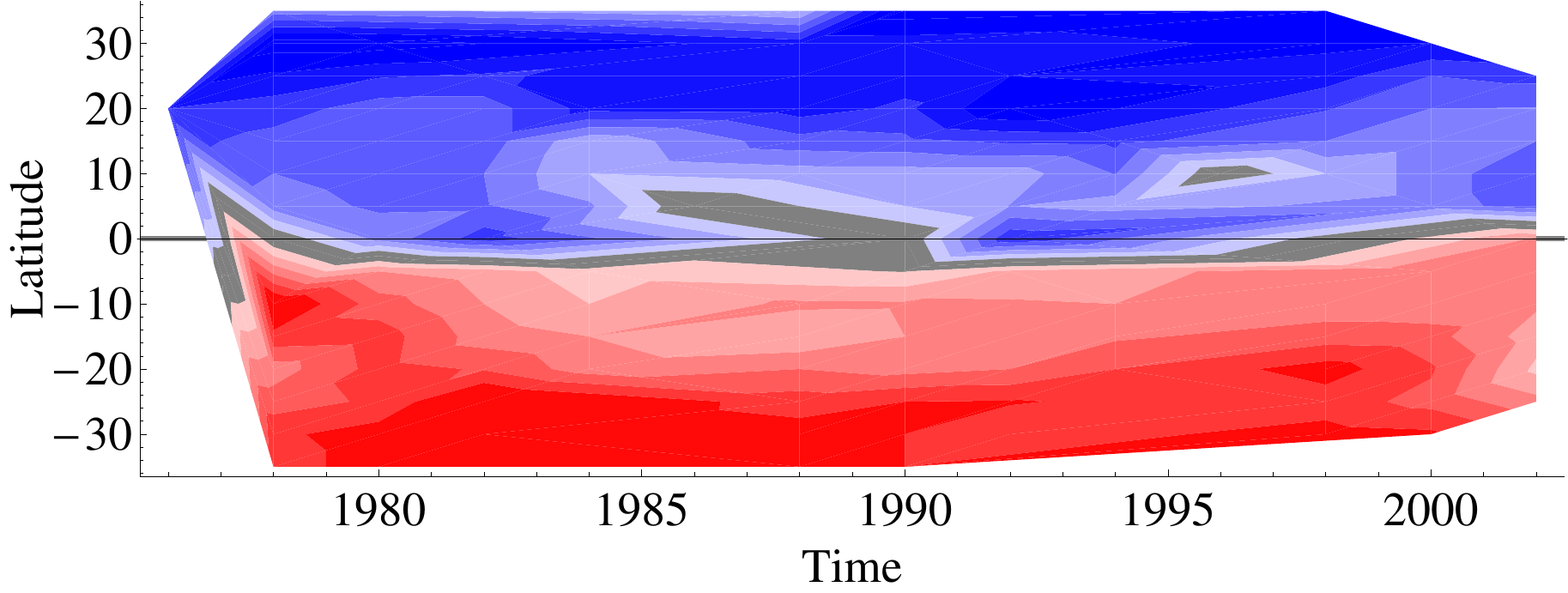}
\caption{Butterfly diagram for tilt alone according to KPVT. 
$B_{\rm min} = 100$ G, areas $S>100$ MSH.
The color notation as in previous figure.}
 \label{Fig9}
\end{figure}

By increasing the lower limit $B_{\rm min}$ to 100 G equatorward 
propagating patterns, connected with cycles, almost disappear. 
In Fig.~\ref{Fig8} and Fig.~\ref{Fig9} we have used  a weighting to increase
the contribution of the largest domains, which include sunspots.
We confirm that the tilt for large bipolar groups at given latitude 
does not vary substantially (Fig.~\ref{Fig9}). 

\begin{figure}
\centering
\includegraphics[width=0.95\linewidth]{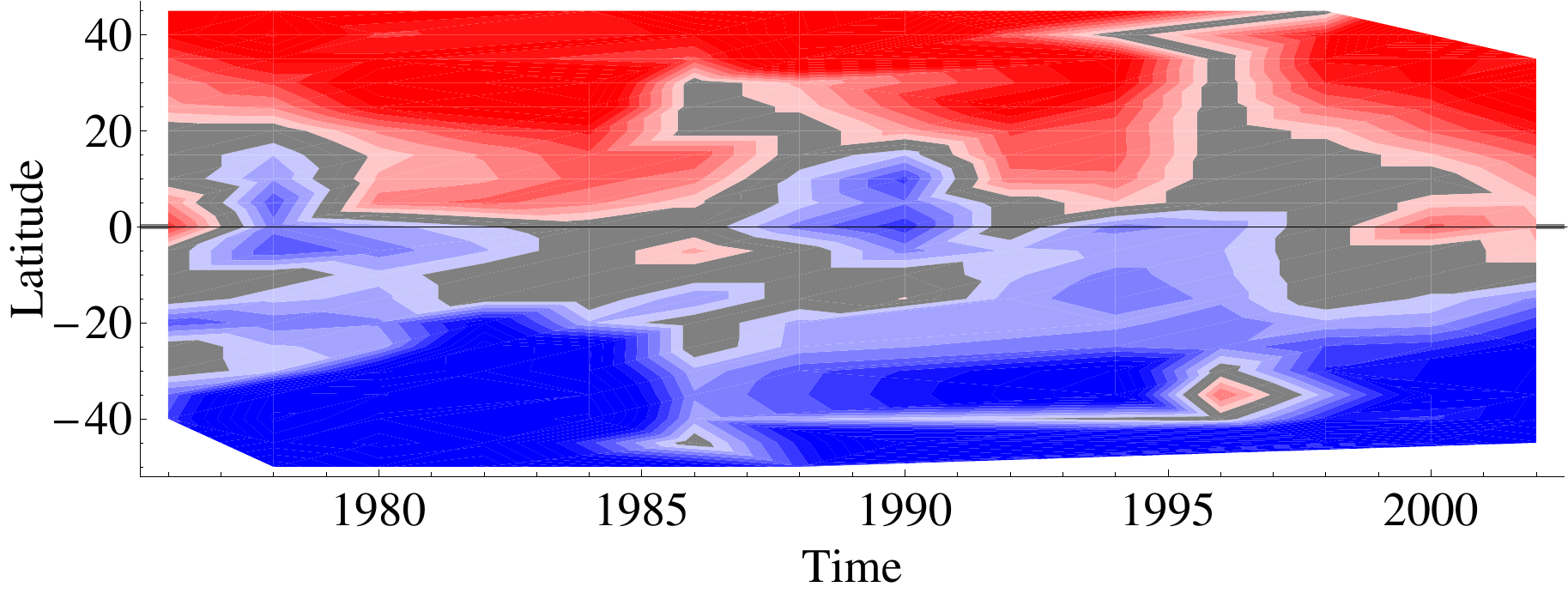}
\caption{Butterfly diagram for tilt alone according to KPVT. 
$B_{\rm min} = 10$ G, areas $50<S<300$ MSH.
The color notation as in previous figure.}
 \label{Fig10}
\end{figure}

The corresponding picture for small bipoles gives, as expected, the inverse color
distribution. In Fig.~\ref{Fig10} we see well defined equatorward 
propagating waves indicating that the small bipoles are 
involved in the solar cycle and give a butterfly diagram similar to 
that obtained from the sunspot data. Note however, that these waves
seem to begin at higher latitudes compared with waves from sunspots.
This could be an indication of the so-called "extended solar cycle".

\section{Discussion}

We have demonstrated that the method for isolation of bipoles 
suggested by Tlatov et al. (2010) provides sufficient 
data to address the time-latitude distribution of the 
tilt angle in detail and to obtain Joy's law for particular 
time-latitude ranges. In this sense our results generalize 
the results obtained by Stenflo \& Kosovichev (2012).

We found that the tilt angle averaged over suitable time-latitude 
bins is distributed antisymmetrically with respect to the solar equator. 
As far as we can follow the latitude distribution, the absolute value 
of the tilt angle is larger for larger latitudes. 

Quite unexpectedly, we obtained that the results depend on the 
size of bipoles included in the investigation, i.e. the large 
bipoles with $S> 300$ MHS (which are mainly sunspot groups) 
and small bipoles with $50 < S < 300$ MHS (which are mainly 
ephemeral regions) give significantly different results concerning 
the tilt angle behaviour. The results for large bipoles look much more significant and easy 
to interpret than those for the small bipoles. In particular, 
the results here agree (at least qualitatively) with the results
obtained by Stenflo \& Kosovichev (2012).  We find that the tilt 
angle is positive in the Northern hemisphere and negative in the
Southern.

Note however that we obtain a smaller inclination factor when the fitting 
the tilt data as being proportional to $\mu = a \sin \pi \theta/180^\circ$. 
We obtained $a = 27\pm 4$  ($2\sigma$ error bar is given)
while Stenflo \& Kosovichev(2012) suggest $a = 32$. We believe that 
the discrepancy between the data analyses reflects the fact that the databases
used are specifically biased by smaller bipoles. 
In other words, the formal statistical uncertainties shown in the figures 
do not take into account a possible bias from the small bipoles. Of course, 
this possible bias constrains recognition of various features found in the 
tilt butterfly diagrams.

{
The results presented on Figures 10-14 are new and they
generalize   the previous findings (see, Tlatov et al, 2010;  
Stenflo \& Kosovichev, 2012) 
giving information about the time-latitude variation of the tilt 
angle for the bipolar regions of different size. 
It was found that the tilt 
angle distribution remains stable showing no substantial variations 
from one cycle to another. For the large bipoles the variations 
in the tilt angle which occur during a given solar cycle mainly reflect 
the propagation of the activity wave to the equator. 
It is found that for the large bipoles the average sign of the tilt
in the given hemisphere is preserved for the separated cycles. This is
different  from what is observed 
in the time-latitude distribution 
for the current helicity (Zhang et al. 2010). }

Solar dynamo theories include a physical link 
between toroidal and poloidal solar magnetic field, which results 
in a non-vanishing tilt. The nature of this link is specific for 
any particular dynamo scenario. In the Parker (1955) scenario it is 
a result of the $\alpha$-effect arising from the mirror-asymmetric solar 
convection, while the Babcock-Leighton scenario (Babcock, 1961, Leighton, 1961)
exploits a direct action of the Coriolis force on the rising magnetic
loop. This can be interpreted as a nonlocal  $\alpha$-effect. 
The direct numerical simulations by Brandenburg (2005) suggest another
possibility. His results suggest that near the solar surface the
$\alpha$-effect can be dominated by effects of the shear, and the tilt 
angle is a result of the latitudinal differential rotation. 
Except for the mechanisms mentioned  above, both mean-field theory and 
direct numerical simulations suggests the existence of additional
induction effects of the turbulence, which could transform the toroidal
magnetic field to poloidal and vice versa. The $\Omega\times J$ effect
(R\"adler, 1969) -- also called the $\delta$-effect -- originates from
the Coriolis force acting on the current, and is induced by the 
turbulent magnetic field linked with the large-scale field 
(see, e.g., Pipin \& Seehafer 2009). Direct numerical
simulations confirm the importance of the $\delta$-effect as well as the
$\alpha^2$ effect for the global dynamo in the convective zone of  a
star (K\"apyl\"a et al 2008, 2009; Schrinner et al 2011, 2012). 

Of course, details of the 
time-latitude distribution of the tilt depend on particular features 
of the model and the scenario as a whole. It looks plausible that 
the tilt data can help to determine which scenario or combinations fits 
the true solar activity better (e.g. Dasi-Espuig (2010) claim that 
the tilt data support the Babcock-Leighton scenario). Our feeling 
is that a final decision here needs extensive numerical modelling 
in light of the tilt data obtained, and so we reserve our opinion here 
and use below the notation $\alpha$ as a common notation for the link 
between toroidal and poloidal magnetic fields in either picture.   

From the viewpoint of dynamo theory, the tilt data can be considered as 
a direct observational determination of the principal drivers in solar 
dynamo, i.e. the $\alpha$-effect (or, more carefully, of a mirror-asymmetric 
driver responsible for conversion of toroidal magnetic field to the 
poloidal). Our results confirm the basic expectations of dynamo 
models concerning properties of $\alpha$. The quantity 
is antisymmetric in respect to the equator, it increases with latitude 
and does not demonstrate pronounced cycle-to-cycle variations. Some variations 
of $\alpha$ within the cycle can be attributed to nonlinear 
suppression of dynamo action.  Note, theoretical
calculations (see, e.g., Fig.1e in Pipin \& Kosovichev 2011) as well as 
direct numerical simulations (e.g., K\"apyl\"a et al. 2006, 2008) 
show that $\alpha$ is positive in the Northern hemisphere in most     
of the convection zone. It changes sign to become negative near the bottom 
of the convection zone.

We can estimate the absolute value of $\alpha$   as follows. 
The typical tilt does not exceed $\mu = 10^\circ$ -- about $10\%$ of 
the extent of the first quadrant, $0^\circ < \mu < 90^\circ$, where the tilt 
of the given sign can, principle, vary. If we identify the life time 
of bipoles with the convection turnover time and suggest that a 
convective vortex performs a full rotation during the turnover time, 
we find that $\alpha \approx 0.1 v$ 
where $v$ is the r.m.s. convective 
velocity. A more accurate estimate requires detailed numerical modelling 
and is constrained by the bias discussed above; however the range of $\alpha$ 
obtained looks consistent with the concepts of dynamo theory.

In any case we can conclude that $\alpha$ is no longer just a theoretical 
concept supported only by theoretical findings, but has now a direct
observational confirmation. Our results do not clarify the physical 
nature of $\alpha$ directly, but further modelling of the dynamo 
based on the data obtained in the framework of other relevant data 
can, in principle, help to decide a relevant option.  
Such modelling however is obviously out of the scope of this paper.

\begin{figure}
\centering
\includegraphics[width=0.9\linewidth]{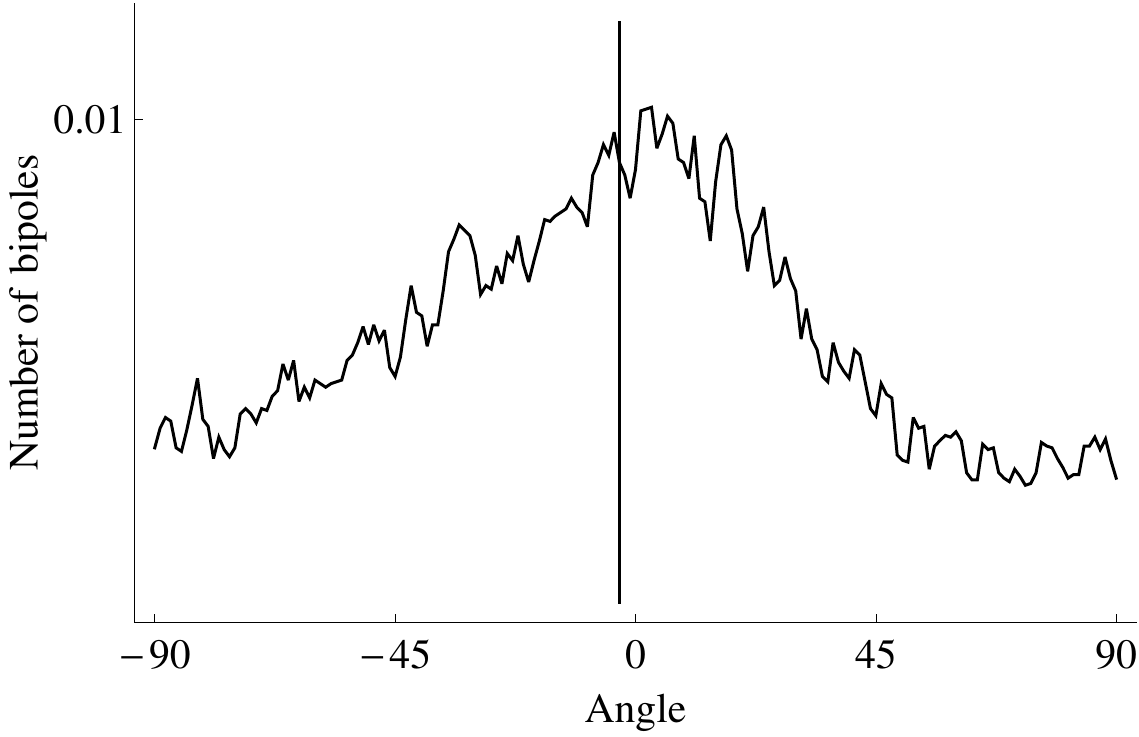}
\caption{Tilt distribution for MDI data (for 1996 - 2011) at 
$12.5^\circ\le\theta\le17.5^\circ$
for areas $50<S<300$ MSH (non-weighted data). Vertical line indicates
median of sample. Graphic is normalized to unit square.}
 \label{dist_s15}
\end{figure}

\begin{figure}
\centering
\includegraphics[width=0.9\linewidth]{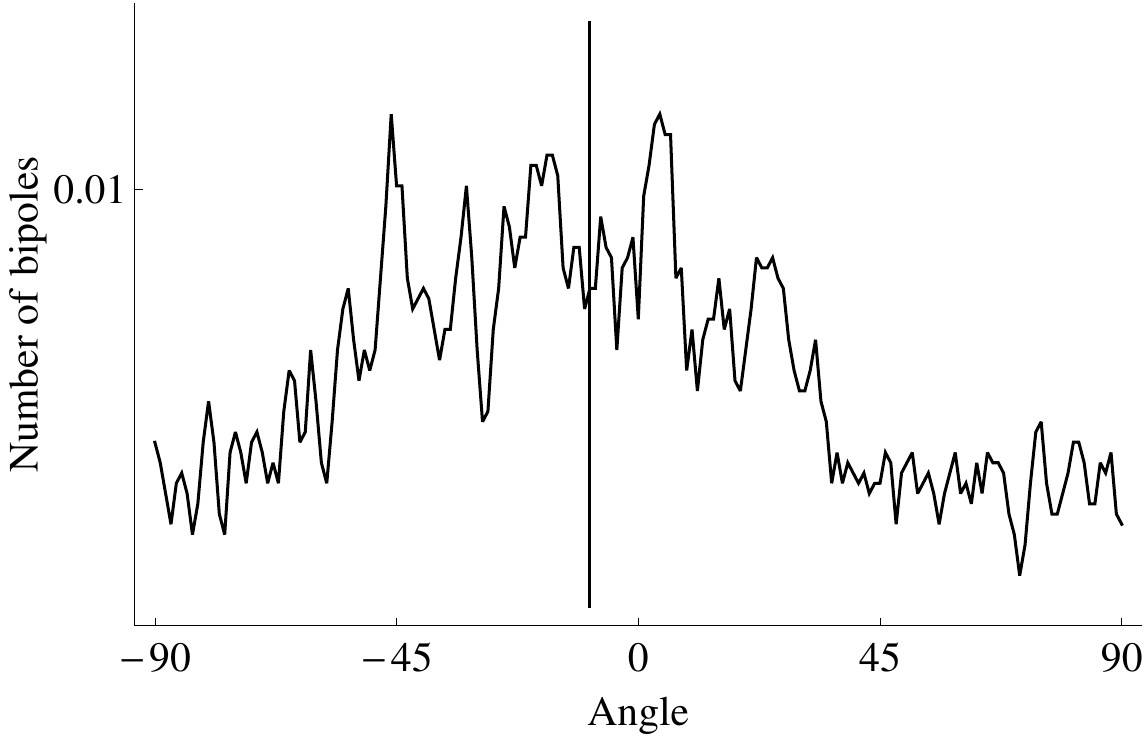}
\caption{Tilt distribution for MDI data (for 1996 - 2011) 
at $32.5^\circ\le\theta\le37.5^\circ$, and areas $50<S<300$ MSH.}
 \label{dist_s35}
\end{figure}

The tilt data associated with small bipoles 
show that these have a sign of tilt 
which is  opposite to the tilt sign derived 
from the data for large bipoles. 
Comparing  the distribution of the tilts of the small bipoles shown in
Fig.s \ref{dist_s15} and \ref{dist_s35} we conclude that the 
origin of the negative tilt  is different at low and high latitudes.

In fact, at low ($<30^\circ$) latitudes the negative tilt is caused not by the
 sign of the maximum of the distribution, but by the more populated tail of the distribution.
We can interpret this as a mixture of two populations.
{ This is different to results by Stenflo \& Kosovichev(2012), who
  found that small bipoles follows to the standard Joy's law being
  oriented in the same direction as the sunspots.} 
One is located in the area of positive angles (standard bipoles) and
defines the maximum of the distribution, whereas another is located in the area
of negative angles and shifts the mean tilt. At higher latitudes ($>30^\circ$) 
the number of standard bipoles with positive tilts (which are supposed
to trace the toroidal magnetic field) decreases and we see the prevalent
component with negative tilt.
{
These results complement the study by Tlatov et al. (2010) who found that the size of population of the oriented small bipoles
  starts to grow at the latitudes larger than 20$^{\circ}$ just after the
  solar maximum having the polar and equatorial branches of activity. It
was claimed that this support an idea about an extended solar
cycle. We can speculate about another possibility.} 

Extending the idea given by Brandenburg (2005), we can assume, 
that at higher latitudes
we observe tracers of poloidal component of the solar magnetic field
that is converted by the $\alpha^2$ mechanism into the toroidal field.
Indeed, such bipoles, being initially oriented in the meridional direction,
can obtain negative tilts from to the action of the Coriolis force.      
For meridionally oriented bipoles the Coriolis force is minimal
near the equator and maximal near the poles, hence the tilt should decrease
towards the poles and we should again  see Joy's law. We can see
such behaviour in Fig.~\ref{Fig2}, where the curve of Joy's law
has a visible bend at $\theta = -30^\circ$. 
Of course, other mechanisms, for instance differential rotation,
can affect the tilt distribution and cause difficulties for the sign determination,
especially at lower latitudes.  

We appreciate that  the criteria for bipole definition and averaging
can be varied to some extent.
These variations affect the tilt distribution of the larger dipoles
more weakly than for the smaller.
{We used median as a robust statistic and Student criteria 
for confidence intervals.
This method confirmed the first estimations in Tlatov et al. 2010.}
However we failed to find a convincing scheme of data 
processing which would give positive tilt for the small bipoles in the Northern
hemisphere. The choice of criteria 
which we used to obtain the  the above results is one of
the most severe in respect to obtaining a negative tilt of small
bipoles in the Northern hemisphere.
{This conclusion is supported by consideration 
of the newest observational data provided by HMI.
}

Our findings can be summarized as follows. For
bipoles  with larger area, the tilt angle is positive
in the Northern hemisphere and negative in the  
Southern, with a latitudinal profile 
proportional to $\cos\theta$. 
{In comparing our results with those by Tlatov et al. (2010) we found that
the bipoles with smaller area are non-homogeneous and 
can be decomposed into two populations.} 
One of them is associated with the bipoles that follow the standard
tilt angle law and contains mostly sunspots at the early stages
of their evolution. Another group is constituted mainly of the
ephemeral regions and has a tilt
angle which is of opposite sign to the tilt of the 
bipoles of larger area. The tilt of bipoles of the second group 
is produced by specific mechanisms,
and some possible explanations have been discussed above.
The ratio between two groups depends on the latitude zone considered.
At low latitudes bipoles of the first group dominate, but
the second group provides a heavy tail to the tilt distribution.
At high latitudes the number of sunspots vanishes 
and the second group dominates. { Extending the results of the
previous studies (see, Tlatov   et al, 2010;  Stenflo \& Kosovichev, 2012; 
and by Tlatov \& Obridko, 2012) we found that the tilt angle distribution 
remains stable showing no substantial variations from one cycle to another.
}

{ Finally, we summarize what is the new message from our paper to dynamo 
theory in comparison to that one from Stenflo \& Kosovichev, 2012.
We focus the attention on the 
variations tilt angle distribution in course of a given cycle to found that 
this distribution remains rather stable 
and deviations from this general distribution are quite small and local. In this 
respect, tilt angle differs substantially from the current helicity which traces 
mirror asymmetry of magnetic field (Zhang et al., 2010).  This simple behaviour 
of the tilt means that the simple models of 
solar dynamo based on primitive parameterization of the regeneration rate of 
poloidal magnetic field from toroidal one and simple phenomenological algebraic 
quenching of this regeneration rate reproduce rather well the true phenomenology. 
Of course, such models do not clarify physical mechanisms underlying the phenomenon however provide
a reasonable pragmatic approach to its description. We believe that the tilt 
angle data presented can be used in order to improve such simplified models for stellar dynamos and historical 
reconstructions of solar activity. In principle, we would be happy to achieve this limited however important goal. 
In fact however we isolated an unusual behaviour of small bipoles. We explain it as a manifestation of regeneration 
of toroidal magnetic field from poloidal one due to the mirror asymmetric effects in solar dynamo. If this 
interpretation will be supported by future investigations, it will give a completely new source of observational 
information concerning governing parameters of solar dynamo.    
}

We stress that an independent verification of the results concerning 
small bipoles by other methods of bipole recognition is highly 
desirable before a dynamo interpretation of this result can be 
robustly supported or rejected. 

\subsection*{Acknowledgements} 
The paper is supported by RFBR under grants 13-02-91158 (AT, EI), 12-02-00614 (AT), 
12-02-31128 (EI), 12-02-00884 (EI), 12-02-00170 (EI, DS, VP). 
We are grateful to J.Stenflo  and D.Moss for a critical reading of the manuscript,
and helpful comments.

\end{document}